\documentstyle[11pt]{article}

\begin{document}
\begin{titlepage}
\begin{flushright}
{\small DE-FG05-92ER40717-5}
\end{flushright}
\vspace{24mm}

\begin{center}
{\LARGE \bf RESOLUTION OF THE $\lambda\Phi^4$ PUZZLE \\
\vspace*{3mm}
            AND A 2 TeV HIGGS BOSON }
\end{center}

\vskip 35 pt plus1pt minus1pt
\centerline{ M. Consoli}
\vskip 4pt plus1pt minus1pt
\centerline{Istituto Nazionale di Fisica Nucleare, Sezione di Catania, Catania,
Italy}
\centerline{Corso Italia 57, 95129 Catania, Italy}
\vskip 10 pt
\centerline{ and}
\vskip 10 pt
\centerline{ P. M. Stevenson}
\vskip 4pt plus1pt minus1pt
\centerline{T. W. Bonner Laboratory, Physics Department }
\centerline{Rice University, Houston, TX 77251, USA}
\vskip 30 pt plus1pt minus 3pt
\centerline{ABSTRACT}
\vskip 8pt plus1pt minus1pt
\baselineskip 15 pt
\par
We argue that massless $(\lambda \Phi^4)_4$ is ``trivial'' without being
entirely trivial.  It has a non-trivial effective potential which leads to
spontaneous symmetry breaking, but the particle excitations above the broken
vacuum are non-interacting.  The key to this picture is the realization that
the constant background field (the mode with zero 4-momentum) renormalizes
differently from the fluctuation field (the finite-momentum modes).  This
picture reconciles rigorous results and lattice calculations with one-loop
and Gaussian-approximation analyses.  Because of ``triviality'', these latter
two methods should be effectively {\it exact}.  Indeed, they yield the same
renormalized effective potential and the same relation $m_h^2 = 8 \pi^2 v^2$
between the particle mass and the physical vacuum expectation value.
This relation predicts a Higgs mass $m_h \sim$ 2 TeV in the standard model.
The non-interacting nature of the scalar sector implies, by the equivalence
theorem, that Higgs and gauge bosons interact only weakly, through their
gauge and Yukawa couplings.

\end{titlepage}
\setcounter{page}{1}
\section{INTRODUCTION}
\setcounter{equation}{0}

  The standard model of electroweak interactions \cite{WS} makes use of
spontaneous symmetry breaking (SSB) to explain the origin of vector-boson
masses \cite{higgs}.  The traditional description relies on an essentially
{\it classical} treatment of a $\lambda \Phi^4$ scalar sector, with
perturbative quantum corrections.  In this picture the Higgs mass $m_h$ is
proportional to the ``renormalized coupling'' $\lambda_R$, so if the Higgs
is heavy ($m_h \ge 0.7$ TeV or so), perturbation theory clearly breaks down.
However, in that case it is usually inferred that longitudinal gauge bosons
would necessarily become strongly interacting at TeV energies \cite{hhg}.

   However, one should really ask if a full {\it quantum} treatment of the
scalar sector can give SSB, and if so what the consequences are.  Here the
whole question is thrown open by the strong evidence that the $\lambda \Phi^4$
theory is ``trivial'' \cite{froh,book,latt,broken}.  Some authors interpret
this to mean that the scalar sector of the standard model can only be an
effective theory, valid only up to some cutoff scale.  Without a cutoff, the
argument goes, there would be no scalar self-interactions, and without scalar
self-interactions there would be no symmetry breaking \cite{call}.  This
view also leads to upper bounds on the Higgs mass \cite{call,neu}.

    However, symmetry breaking is not incompatible with ``triviality''.
One could have a non-zero vacuum expectation value for the field, yet
find only non-interacting, free-particle excitations above the vacuum.
We argue that this is what happens in pure $\lambda \Phi^4$ theory.
Our picture \cite{con,new,iban,u1} reconciles the evidence for
``triviality'' with the evidence for a non-trivial effective potential.

    A non-trivial effective potential with SSB \cite{cian} emerged naturally
from an analysis of the Gaussian effective potential \cite{return,cast}
of $\lambda \Phi^4$ theory.  More recently it was realized that the same
results emerge from the one-loop effective potential \cite{bran}.  The key
features are ``asymptotic freedom'' ({\it i.e.,} a flow of the bare coupling
constant $\lambda_B$, as a function of the lattice spacing $a$, in which
$\lambda \to 0$ as $a \to 0$, such that a macroscopic correlation length is
obtained) and {\it masslessness} in the symmetric phase.

   Initially, it was thought \cite{return} that the Gaussian-approximation
results implied a fully non-trivial, interacting theory.  However, since the
effective potential contains information only about zero-momentum Green's
functions, it does not by itself provide information about the existence or
otherwise of a momentum-dependent renormalized coupling ``$\lambda_R(Q^2)$.''
Consequently, these results do not necessarily imply the existence of
non-trivial interactions in the ``shifted'' theory.
(In fact, the finite-temperature generalization \cite{hajj} implies that the
shifted field is non-interacting, and apparent paradoxes in the extension
to the effective action, variously interpreted \cite{ea}, may be seen as
pointing to the same conclusion.)

   In this paper we elaborate the following picture for pure $\lambda \Phi^4$
theory \cite{con,new,fnote}:  Writing the field $\Phi(x)$ as $\phi+h(x)$, we
expect the fluctuation field $h(x)$ to be non-interacting, but its mass will
be proportional to the background constant field $\phi$.  Thus, its vacuum
energy is $\phi$ dependent and this, together with the $\lambda \phi^4$ term,
produces a non-trivial one-loop effective potential, which should be
{\it exact}, because there are no $h$-particle interactions.  The
renormalization requires the bare coupling constant to go to zero in the
continuum limit, and involves an infinite re-scaling of the $\phi$ field.
The $h(x)$ field, however, is not re-scaled, and this provides a simple way
to understand why it is non-interacting.  Most of the paper is concerned with
pure $\lambda \Phi^4$ theory, but we consider the implications for the
electroweak theory in the final section.

\section {``TRIVIALITY'' AND SSB}
\setcounter{equation}{0}

    Analytical and numerical studies of $(\lambda\Phi^4)_4$ theory
\cite{froh,book,latt,broken} have overwhelmingly been driven to the conclusion
that it is a ``generalized free field theory.''  This means that all
renormalized Green's functions of the continuum theory are expressible in
terms of the first two moments of a Gaussian distribution \cite{glimm}:
\begin{equation}
\label{e1}
    \tau(x)=v,
\end{equation}
\begin{equation}
\label{e2}
    \tau(x,y)=v^2 + G(x-y),
\end{equation}
so that
\begin{equation}
\label{e3}
    \tau(x,y,z)= v^3+ v(G(x-y)+G(x-z)+G(y-z)),
\end{equation}
\begin{equation}
\label{e4}
   \tau(x,y,z,w)=
v^4+ v^2(G(x-y)+{\rm perm.})+G(x-y)G(z-w) + {\rm perm.},
\end{equation}
and so on.  Here, $v$ is a constant (since we assume that translational
invariance is not broken), and $G(x-y)$ is just a free propagator with some
mass $m_h$ and residue $Z_h=1$, since it must satisfy a K\"{a}llen-Lehmann
representation with spectral function $\delta(s-m^2_h)$.  The index ``$h$''
in $Z_h$ and $m_h$ refers to the shifted field $h(x)$ introduced by means of
a suitably de-singularized, renormalized field operator $\Phi_R(x)$, such that
$\langle \Phi_R(x) \rangle = v$ and $h(x) \equiv \Phi_R(x) - v$.  The above
equations imply that all {\it connected} three- and higher-point Green's
functions of the $h(x)$ field vanish; {\it i.e.}, ``triviality''.

However, although the generalized free-field structure implies a trivially
free shifted theory it does not forbid a non-zero value of $v$.  Indeed,
there are explicit studies of ``triviality'' in the broken phase ($v \ne 0$)
\cite{broken}.  This suggests that it is possible for the theory to have
a non-trivial effective potential $V_{{\rm eff}}$ with SSB minima.
Indeed, the lattice calculations of Huang {\it et al} \cite{huang} do find
a non-trivial effective potential, even while finding no sign of particle
interactions: they conclude that the theory cannot be {\it entirely}
trivial \cite{huang,huang2}.  The effective potential, when expanded
about the origin, generates the zero-momentum Green's functions of the
symmetric phase \cite{CW}.  Thus, for the effective potential to be
non-trivial and give SSB, there has to be non-triviality in the
zero-momentum-mode sector of the symmetric phase, which somehow acts to
de-stabilize the symmetric vacuum.

This immediately suggests that we concentrate on {\it massless}
$\lambda\Phi^4$ theories, for which zero-momentum ($p^{\mu}=0$) represents
a physical, on-shell point.  For a massless theory, a non-trivial effective
potential would imply a non-trivial scattering matrix.  There is, in fact,
good evidence that massless $\lambda\Phi^4$ theory {\it is} non-trivial in
this sense \cite{ped}.
[In Ref. \cite{ped} the conformal symmetry of the massless $\lambda\Phi^4$
theory is exploited, allowing a mapping of the Minkowski space into the
Einstein universe.  Asymptotic states are restricted to those which carry
``Einstein quantum numbers''.  In Minkowski space these are collective
states containing an indefinite number of massless particles.  As in the
well-known Bloch-Nordsieck procedure in QED, such states are needed because
of the serious infrared problems.  The effective potential also involves a
sum of diagrams with an arbitrarily large number of external legs, with
infrared divergences cancelling only in the sum \cite{CW}.]

It is quite natural to expect that any non-trivial dynamics of the
zero-momentum mode would induce SSB:  The classical potential of the
massless theory is very flat at the origin, and ripe for instability.
Moreover, this would be the simplest way for the theory to escape its
infrared difficulties.  Since the massless theory contains no intrinsic scale,
the physical scale, $v$ (with $m_h$ proportional to $v$), must be spontaneously
generated by ``dimensional transmutation'' from a theory with no intrinsic
scale in its symmetric phase.  This is exactly the philosophy of the classic
paper of Coleman and Weinberg \cite{CW}.

``Dimensional transmutation'' requires ultraviolet divergences to produce
a non-zero Callan-Symanzik $\beta$ function.  A familiar example is the
origin of the $\Lambda$ scale in QCD.  However, in that case one can
obtain the $\beta$ function perturbatively, from the momentum dependence of
the renormalized coupling constant.  In $(\lambda \Phi^4)_4$ theory that
approach is doomed to failure since ``triviality'' means that any
``renormalized coupling constant'', defined from the 4-point function at
non-zero external momenta, must vanish.  (It is not surprising, then, that
the perturbatively defined $\beta$ function is positive and has no fixed
point.  This implies an unphysical Landau pole at high energies -- which can
only be avoided by requiring the renormalized coupling at finite energies to
vanish as the ultraviolet regulator is removed.  In this sense, perturbation
theory itself signals ``triviality'' \cite{call}.)  To extract a more
meaningful $\beta$ function one should start from a quantity that will be
finite, and {\it non-vanishing} in the infinite-cutoff limit.  This is not
true for the 4-point function.

However, in our picture, $V_{{\rm eff}}$ is non-trivial and one may extract
from it a nonperturbatively defined $\beta$ function that characterizes the
dependence of the {\it bare} coupling constant on the ultraviolet regulator.
For the picture to be consistent we would expect this $\beta$ function to be
negative, giving asymptotic freedom ($\lambda_B \to 0$ in the continuum limit),
since it seems \cite{froh} that asymptotic freedom is the only possibility to
avoid ``entire triviality''.

\section{THE EFFECTIVE POTENTIAL}
\setcounter{equation}{0}

To find $\beta$ from $V_{{\rm eff}}$ we follow the usual Renormalization-Group
(RG) procedure.  First, we introduce a generic ultraviolet regulator ``$r$''
(which may be a cutoff $\Lambda$, or an inverse lattice spacing $1/a$, or one
may identify $\ln r^2$ with ${{2}\over{4-d}}$
in dimensional regularization, etc.), so that the effective potential depends
on $r$, and on the bare classical field $\phi_B$ and on the bare mass and
coupling parameters:
\begin{equation}
\label{e5}
  V_{{\rm eff}}= V_{{\rm eff}}(\phi_B,\lambda_B,m^2_B,r).
\end{equation}
The bare mass is treated as a counterterm for the quantum theory such that,
in any approximation and in any regularization scheme, the condition
\cite{CW}
\begin{equation}
\label{e6}
\left. {{\partial^2 V_{{\rm eff}}}\over{\partial \phi^2_B}}
\right|_{\phi_B=0} = 0
\end{equation}
is satisfied.  This ensures that the theory has no intrinsic scale in its
symmetric phase. (In dimensional regularization this simply requires $m_B=0$
and the classical scale invariance of the bare Lagrangian is manifest.)
After implementing the masslessness condition, (\ref{e6}), the effective
potential depends only on $r$ itself and on the bare parameters
$\phi_B=\phi_B(r)$ and $\lambda_B=\lambda_B(r)$ whose flow, in the limit
$r\to \infty$, is dictated by the requirement of RG invariance; that is:
\begin{equation}
\label{e7}
\lim_{r\to\infty}~ \left \{ r {{\partial}\over{\partial r}} + \beta
{{\partial}\over{\partial\lambda_B}} -
\gamma \phi_B {{\partial}\over{\partial\phi_B}}
\right \} V_{{\rm eff}}(\phi_B , \lambda_B , r ) = 0,
\end{equation}
where $\beta \equiv r {{d \lambda_B}\over{d r}}$ and
$\gamma \equiv - {{r}\over{\phi_B}} {{d \phi_B}\over{d r}}$.
This anomalous dimension $\gamma$ should not be confused with the
more conventional quantity associated with the $p^2$-dependence of the
two-point function.  (That quantity would be associated with the cutoff
dependence of $Z_h$, and must vanish as $r \to \infty$ since $Z_h \to 1$.)
Here, $\phi_B$ is a constant background field, and has no dependence on $p^2$.

One can easily disentangle $\beta$ from $\gamma$ as follows \cite{cast}.
If the effective potential has its non-trivial minimum at a particular
$\phi_B = v_B$; i.e.,
\begin{equation}
\label{e8}
\left. {{\partial V_{{\rm eff}}}\over{\partial \phi_B}}
\right|_{\phi_B = v_B} = 0,
\end{equation}
where $v_B$ will depend on $r$ and $\lambda_B$, then the vacuum energy density:
\begin{equation}
\label{e9}
  W(\lambda_B,r)=
V_{{\rm eff}}(v_B{\scriptstyle{(r,\lambda_B)}},\lambda_B,r)
\end{equation}
will satisfy (from Eq. (\ref{e7}))
\begin{equation}
\label{e10}
\lim_{r\to\infty}~ \left \{ r {{\partial}\over{\partial r}} + \beta
{{\partial}\over{\partial\lambda_B}}
\right \} W( \lambda_B , r ) = 0.
\end{equation}
This equation will determine $\beta$ and one can then go back to (\ref{e7})
to extract $\gamma$.

   Acceptance of ``triviality'' in the sense of Eqs. (\ref{e1}--\ref{e4})
means that this procedure represents the only hope for obtaining a
not-entirely-trivial continuum limit.  (It also means that we may expect a
contradiction if we try to use an approximation to the effective potential
that is inherently inconsistent with generalized-free-field behaviour.
We shall return to this important point in Sect. 6.)

\section{THE ONE-LOOP EFFECTIVE POTENTIAL}
\setcounter{equation}{0}

    Now let us see what happens in the simplest approximation scheme.
We first write the bare field as
\begin{equation}
\Phi_B(x) = \phi_B + h_B(x),
\end{equation}
where we have separated out the zero-momentum component
$\phi_B = {\rm constant}$ (demanding, in order for the separation to be
unambiguous, that the field $h_B(x)$ has no Fourier projection on to
$p^{\mu}=0$).  Consider the approximation in which the field $h_B(x)$ is
allowed to interact to all orders in $\lambda_B$ with $\phi_B$ but has no
non-linear interaction with itself.  To this level of approximation $h_B(x)$
is just a free field whose mass-squared is
\begin{equation}
\omega^2{\scriptstyle (\phi_B)} \equiv {{1}\over{2}} \lambda_B \phi^2_B.
\end{equation}
However, through its zero-point energy, it will produce a non-trivial
contribution to the effective potential.  To compute this we take the
shifted, linearized Hamiltonian:
\begin{equation}
   H_o(\phi_B)=\int {\rm d}^3\vec x \left [{{\lambda_B}\over{4!}}\phi^4_B
+{{1}\over{2}}\Pi^2_h+{{1}\over{2}}(\vec\nabla h)^2+
{{1}\over{2}} \omega^2{\scriptstyle (\phi_B)} \, h^2 \right ] ,
\end{equation}
and find its lowest eigenvalue, as a function of $\phi_B$.  This gives the
effective potential times a volume factor.  Dropping a $\phi_B$-independent
constant term, this gives:
\begin{eqnarray}
\label{ol}
 V^{{\rm 1-loop}}(\phi_B) & = & \frac{\lambda_B}{4!} \phi^4_B +
\frac{\omega^4{\scriptstyle (\phi_B)}}{64\pi^2}
\left( \ln \frac{\omega^2{\scriptstyle (\phi_B)} }{\Lambda^2} -
\frac{1}{2} \right),
\\
& = & {{\lambda_B}\over{24}} \phi^4_B +
{{\lambda^2_B\phi^4_B}\over{256\pi^2}} \left( \ln {{\lambda_B\phi^2_B}\over
{2\Lambda^2}}
 - {{1}\over{2}} \right),
\end{eqnarray}
where we use an ultraviolet cutoff $\Lambda$, as in Ref. \cite{CW}.

   This is, of course, the familiar ``one-loop effective potential''
\cite{CW,onel,cjt}.  In Ref. \cite{CW} it was obtained, using Feynman
diagrams of the {\it unshifted} theory, from the sum of all one-loop 1PI
graphs with external lines $\phi_B$ and {\it massless} propagators running
round the loop.  (This diagrammatic interpretation links the effective
potential with the dynamics of the underlying massless theory, as discussed
earlier.)  Although we use the traditional name, ``one-loop'', for this
approximation, we stress that it is the above linearization procedure, and
not the loop expansion, that is our rationale for it.

   Taking this effective potential and following the well-defined procedure
described in Eqs. (\ref{e7}--\ref{e10}) above, one finds straightforwardly
\cite{bran,con}:
\begin{equation}
\label{e12}
 \beta= - \, {{3\lambda^2_B}\over{16\pi^2}},
\end{equation}
\begin{equation}
\label{e13}
 \gamma = - \, {{3\lambda_B}\over{32\pi^2}}
= {{\beta} \over{2\lambda_B}}.
\end{equation}
This result is crucial, and the reader is urged to check it for him- or
her-self.

  Thus, this $\beta$ is indeed negative.  These RG functions allow the limit
$\Lambda\to\infty$ to be taken such that the vacuum energy density
$W =- m^4_h/(128\pi^2)$ is finite and RG invariant.  Hence, the particle mass:
\begin{equation}
\label{e14}
 m^2_h \equiv \omega^2{\scriptstyle (v_B)}  =
\frac{1}{2} \lambda_B v^2_B =
\Lambda^2~\exp (-{{32\pi^2}\over{3\lambda_B}}),
\end{equation}
is finite and RG invariant.  When re-arranged, this equation shows the
asymptotic-freedom property explicitly:
\begin{equation}
\label{e15}
 \lambda_B = {{32\pi^2}\over{3}} {{1}\over{\ln (\Lambda^2/m_h^2)}}.
\end{equation}
Eliminating $\Lambda$ in favour of $v_B$, $V^{{\rm1-loop}}$ can be
expressed as:
\begin{equation}
\label{e16}
 V^{{\rm1-loop}}(\phi_B) =
{{\lambda^2_B\phi^4_B}\over{256\pi^2}} \left( \ln {{\phi^2_B}
\over{v^2_B}}
 - {{1}\over{2}} \right).
\end{equation}

To make the finiteness and RG invariance of $V^{{\rm 1-loop}}$ manifest
we may re-express it in terms of a renormalized field
$\phi_R = Z^{-{{1}\over{2}} }_{\phi} \phi_B$, where an infinite $Z_{\phi}$
absorbs the cutoff dependence.  This $Z_{\phi}$ must satisfy
$ \gamma = - {{1}\over{2}} \Lambda~ {{d}\over{d\Lambda}} \ln Z_{\phi} $
where $\gamma$ is the anomalous dimension of Eqs. (\ref{e7}, \ref{e13}).
The absolute normalization of $Z_{\phi}$ is fixed by requiring the physical
mass $m^2_h$ of the fluctuation field $h_R(x)=h_B(x)$ to be equal to the
second derivative, at the minimum, of $V^{{\rm 1-loop}}$ with respect to
$\phi_R$.  (We shall return to justify this condition in a moment.)

As shown in Refs. \cite{con,iban,new} one obtains
\begin{equation}
\label{e17}
   Z^{{\rm 1-loop}}_{\phi}={{16\pi^2}\over{\lambda_B}} =
{{3}\over{2}} \ln (\Lambda^2/m_h^2),
\end{equation}
so that (with $v \equiv v_R = Z_{\phi}^{-{{1}\over{2}}} v_B$)
\begin{equation}
\label{e18}
 V^{{\rm 1-loop}} =
{}~\pi^2 \phi^4_R \left( \ln {{\phi^2_R}\over{ v^2}}
 - {{1}\over{2}} \right),
\end{equation}
and
\begin{equation}
\label{e19}
m^2_h~ = ~8\pi^2 v^2,
\end{equation}
as advertised earlier.

\section{THE FIELD RE-SCALING}
\setcounter{equation}{0}

  We now want to discuss two apparently unconventional aspects of this
analysis.  Firstly, it is crucial to this picture that the $Z^{1/2}_{\phi}$
re-scaling of the constant background field $\phi_B$ is quite distinct from
the $Z^{1/2}_h = 1$ re-scaling of the fluctuation field $h(x)$.  We can, in
fact, express this as a single, overall re-scaling of the whole field,
provided that we use a momentum-dependent $Z^{1/2}(p)$:
\begin{equation}
\label{e20}
Z^{ {{1}\over{2}} }(p) = Z^{ {{1}\over{2}} }_{\phi}{\cal P} +
Z^{ {{1}\over{2}} }_h \overline{{\cal P}},
\end{equation}
where
\begin{equation}
\label{e21}
 {\cal P} \equiv {{\bar\delta^4(p)}\over{\bar\delta^4(0)}}
{}~~~~~~~{\rm and}~~~~~~~\overline{{\cal P}} = 1 - {\cal P}
\end{equation}
are orthogonal projections (${\cal P}^2 = {\cal P}$,
$\overline{{\cal P}}^2=\overline{{\cal P}}$, ${\cal P}\overline{{\cal P}}=0$)
which select and remove the $p^{\mu}=0$ mode, respectively.  [Here
$\bar\delta^4(p) \equiv (2\pi)^4 \delta^4(p)$, and $\bar\delta^4(0)$ has the
usual interpretation as the spacetime volume.]  Note that ``$p^{\mu}=0$''
is a Lorentz-invariant statement so our momentum-dependent $Z(p)$ does not
violate any sacred principles.  In terms of the Fourier transform of the field
operators it works as follows:
\begin{equation}
Z^{ {{1}\over{2}} }(p) \tilde\Phi_R(p) =
\left( Z^{ {{1}\over{2}} }_{\phi}{\cal P} +
Z^{ {{1}\over{2}} }_h \overline{{\cal P}} \right)
\left( \phi_R\bar\delta^4(p) + \tilde h_R(p) \right)
\nonumber
\end{equation}
\begin{equation}
\label{e22}
 = Z^{ {{1}\over{2}} }_{\phi}\phi_R\bar\delta^4(p) +
Z^{ {{1}\over{2}} }_h \tilde h_R(p) = \phi_B\bar\delta^4(p)+\tilde h_B(p)
= \tilde\Phi_B(p).
\end{equation}
(Note that ${\cal P} \tilde h(p) = 0$ by definition.)  It is easy to check
that $Z^{-1/2}(p)$, $Z(p)=(Z^{1/2}(p))^2$, etc.,  work properly.  For
consistency, the renormalized momentum-space propagator of the complete
$\tilde\Phi_R(p)$ field must be written as:
\begin{equation}
\label{e23}
   \phi^2_R\bar\delta^4(p) +
{{\overline{{\cal P}}}\over{p^2-\omega^2{\scriptstyle(\phi_R)}}},
\end{equation}
with $\omega^2{\scriptstyle(\phi_R)} = 8 \pi^2 \phi_R^2$.
(At $\phi_R = v$ this corresponds to the Fourier transform of Eq. (\ref{e2}).)
The $\overline{{\cal P}}$ projection makes no difference {\it except} in
the symmetric phase (where $\omega^2$ and $\phi_R^2$ vanish): the propagator is
then formally $\overline{{\cal P}}/p^2$, not the free, massless propagator
$1/p^2$.  This form may allow not-entirely-free behaviour, while still being
compatible with the constraints imposed by scale and conformal invariance
(see Ref. \cite{gross}).

Secondly, recall that we fixed the absolute normalization of $Z_{\phi}$ by
requiring the physical mass $m^2_h = {{1}\over{2}} \lambda_B v_B^2$ to be
equal to the second derivative of $V^{{\rm 1-loop}}(\phi_R)$ at its minimum.
This arises from the well-known connection between the derivatives of the
effective potential and the zero-momentum limit of the 1PI Green's functions
\cite{CW}.  Here, in the SSB vacuum, the renormalized inverse propagator of
the $h$ field is just $p^2 - m_h^2$, which tends to $ - m_h^2$ as
$p^{\mu} \to 0$.  This must agree with (minus) the second derivative of
$V_{{\rm eff}}(\phi_R)$ at the vacuum.

We can re-state the argument in more physical terms: For self-consistency
-- especially if we believe the $h$ field to be truly a free field --
the effective potential near its minimum should look like a free-field
potential ${{1}\over{2}} m_h^2 (\phi_R - v)^2$ with the same mass as in the
propagator.  (This is extremely intuitive if one first thinks about the
0+1 dimensional, quantum mechanical case.)  Furthermore, because fluctuations
of $h(x)$ that are finite on the scale of $\phi_B$ are only infinitesimal
on the scale of $\phi_R$, it is quite reasonable that the $h(x)$ field is only
sensitive to the quadratic shape of the renormalized $V_{{\rm eff}}$ in the
infinitesimal neighbourhood of the vacuum.  Therefore, we can understand why
$h(x)$ behaves as a free field, and thus close the circle of our arguments.

    In fact, we may show directly that $h(x)$ is non-interacting.  The
bare 3-point and 4-point vertices of the shifted theory are proportional
to $\lambda_B \phi_B$ and $\lambda_B$, respectively.  These are both
infinitesimal, of order $\sqrt{\epsilon}$ and $\epsilon$, respectively
(where $\epsilon$ is $1/\ln \Lambda$ in cutoff regularization, or $4-d$ in
dimensional regularization).  Any diagram with $T$ 3-point vertices,
$F$ 4-point vertices, and $L$ loops can, at most, be of order
$(\lambda_B \phi_B)^T (\lambda_B)^F (1/\epsilon)^L = \epsilon^{T/2+F-L}$.
But it is an identity that $T/2+F-L = n/2-1$, where $n$ is the number of
external legs.  Thus, all diagrams contributing to the $n$-point function
are suppressed by a factor of at least $\epsilon^{n/2-1}$, and hence
vanish for $n \ge 3$ in the continuum limit.  Thus, there are no $h$
self-interactions, as a direct consequence of the fact that $h(x)$, unlike
the constant $\phi$ field, has no infinite re-scaling that can compensate
for the infinitesimal strength of $\lambda_B$.  Thus, our results confirm our
initial expectation that $\lambda\Phi^4$ is a generalized free field theory.

\section{BEYOND ONE-LOOP: EXACTNESS CONJECTURE}
\setcounter{equation}{0}

  Now, if the $h(x)$ field does not self-interact, then the effective
action:
\begin{equation}
\label{e24}
\Gamma[\Phi_R]=~\int d^4x~
 \left \{  {{1}\over{2}} (\partial h_R)^2- {{1}\over{2}}(8\pi^2\phi^2_R) h^2_R
- \pi^2 \phi^4_R \left( \ln {{\phi^2_R}\over{ v^2}}
 - {{1}\over{2}} \right)
 \right \},
\end{equation}
that embodies our results (\ref{e18}, \ref{e19}), ought to be {\it exact}
\cite{fnotecon}.  The shifted field $h(x)$ has made a non-trivial contribution
to $V_{{\rm eff}}$ through its zero-point energy, but any further
modifications would have to be due to its self-interactions -- and physically
it has none!

   This conclusion would not be immediately obvious diagrammatically,
because there are indeed vacuum diagrams at all orders that give
$1/\epsilon$ and finite contributions, as our argument above (for $n=0$)
shows.  Similarly, for $n=2$, we see that there are diagrams giving finite
contributions to the propagator in all orders.  However, since the $h(x)$
field has no {\it physical} interactions, the only reasonable expectation
is that the apparent ``higher-order corrections'' to $V_{{\rm eff}}$ and
to $m_h^2$ either cancel or can be re-absorbed into the renormalization
constants so as to leave the physical results (\ref{e18}, \ref{e19})
unchanged.

  This conjecture is supported by the observation \cite{con,new} that
exactly the same physical results (\ref{e18}, \ref{e19}) for $V_{{\rm eff}}$
and $m_h^2$ are obtained in the Gaussian approximation.  Discarding terms
that will vanish in the limit $\Lambda\to \infty$, the Gaussian effective
potential for the massless case can be expressed as \cite{con,new}:
\begin{equation}
V^{{\rm Gauss}}=\frac{\lambda_G}{4!} \phi^4_B +
\frac{\Omega^4{\scriptstyle (\phi_B)}}{64\pi^2}
\left( \ln \frac{\Omega^2{\scriptstyle (\phi_B)}}{\Lambda^2} -
\frac{1}{2} \right),
\end{equation}
with
$\Omega^2{\scriptstyle (\phi_B)} = \frac{1}{2} \lambda_G\phi^2_B$ and
$\lambda_G={{2}\over{3}}\lambda_B$.  This has the same structure as the
one-loop result, Eq. (\ref{ol}).  After eliminating $\Lambda$ in favour
of $v_B$, and then re-scaling the bare vacuum field by
$Z^{{\rm Gauss}}_{\phi}={{16\pi^2}\over{\lambda_G}} =
{{24\pi^2}\over{\lambda_B}}$, the two approximations are seen to be
physically equivalent.  There are differences by factors of 2/3 in the
unobservable quantities $\lambda_B$ and $Z_{\phi}$, but these differences
cancel out in the physically meaningful results (\ref{e18}, \ref{e19})
\cite{con,new}.

Notice that both the 1-loop and Gaussian approximations have a
variational character: the one-loop potential is the ground-state energy of
that part of the Hamiltonian which is quadratic in the shifted field, while
the Gaussian approximation corresponds to minimizing the expectation value of
the full Hamiltonian in the subspace of Gaussian wavefunctionals.
The Gaussian approximation can also be described as a re-summation of an
infinite subset, a convergent subseries, of `daisy' and `superdaisy' diagrams
\cite{dj,bg}.

Other truncations of the diagrammatic series for the effective potential,
if they do not correspond to any variational procedure, and hence do not
enjoy a stability property, may give rise to spurious difficulties.
Apparent contradictions will inevitably arise if one tries to use an
approximation method that is inherently incompatible with
generalized-free-field behaviour, Eqs. (\ref{e1}--\ref{e4}).  This is the
case with the usual loop expansion beyond one loop.

  In other words, one should not consider the $\beta$ and $\gamma$ functions
in Eqs. (\ref{e12}, \ref{e13}) as the first terms of a power-series expansion
in $\lambda_B$ which can be systematically improved order by order in the
loop expansion for the effective potential.  Rather, the form of Eqs.
(\ref{e12}, \ref{e13}) will arise in any approximation scheme, no matter how
sophisticated, that allows generalized free-field behaviour to be a possible
solution.  The 2-loop, 3-loop, ... approximations to the effective potential,
on the other hand, have a built-in incompatibility with generalized free-field
behaviour (think of the spurious $Z_h \neq 1$ effect, starting at the 2-loop
level, for instance) and represent merely a {\it perturbative} construction.

  This point can best be understood in terms of the effective potential for
composite operators introduced by Cornwall, Jackiw and Tomboulis \cite{cjt}
(CJT), which is based on the rigorous definition of the effective potential
through the exact relation
\begin{equation}
\label{e25}
\int \! d^3x \, V_{{\rm eff}}(\phi) = E[\phi, G_o(\phi)],
\end{equation}
where $E[\phi, G] = \min \langle \Psi|H|\Psi \rangle$, minimized over all
normalized states $|\Psi \rangle$,  subject to the conditions
$\langle \Psi|\Phi|\Psi \rangle = \phi$
and $\langle \Psi|\Phi(\vec x,t)\Phi( \vec y,t)|\Psi \rangle =
\phi^2 + G(\vec x, \vec y)$,
and where $G_o(\phi)$ is obtained from
\begin{equation}
\label{e26}
\left. {{\delta E}\over{\delta G(\vec x,\vec y)}}\right|_{G=G_o(\phi)}=0.
\end{equation}
Equation (\ref{e26}) can be solved exactly for $G_o$ in the subspace of
Gaussian states and in that case Eq. (\ref{e25}) leads to the Gaussian
effective potential.  The problem can also be approached diagrammatically,
using CJT's result that $E[\phi,G]$ has a manifestly covariant expansion
containing the one-loop term and the series of 2-particle-irreducible vacuum
graphs with propagator $G$ and vertices given by the shifted interaction
Lagrangian.  However, in an $n$-loop approximation to $E[\phi,G]$ ($n \ge 2$),
one cannot exactly solve the resulting optimization equation, (\ref{e26}).
One can only solve it in a {\it perturbative} sense.  Since this does not
provide a true stationary solution for $G$ what one obtains is really not an
effective potential, except in a perturbative sense.  Consequently, one should
not attempt to apply the RG equation, (\ref{e7}), to the 2-loop, 3-loop, ...
approximations to the effective potential.  That would be similar to
trying to define $\beta$ through the perturbative 4-point function at
non-zero external momenta and, as we explained in Sect. 2, that cannot produce
a consistent continuum limit.

  It is possible, in principle, to consider truncations of the CJT
construction that improve upon the one-loop or Gaussian approximations, yet
still allow the resulting optimization equation, (\ref{e26}), to be solved
exactly.   For example, one can consider post-Gaussian variational
calculations (either Hamiltonian \cite{pr} or covariant \cite{ipr}).  In
such a calculation, however, we would expect the optimal $G$ to reduce to
a free propagator, and our equations (\ref{e18}, \ref{e19}) to be unmodified
in the continuum limit.  Any other result would provide strong evidence that
$(\lambda\Phi^4)_4$ is not, in fact, a generalized free field theory.
Therefore, if ``triviality'' is true, as we believe, then Eqs.
(\ref{e18}, \ref{e19}) should be considered {\it exact}.

\section{FINITE TEMPERATURE}
\setcounter{equation}{0}

     We briefly discuss finite-temperature effects in order to make two
points:  (i) our $\lambda \Phi^4$ theory, even though it has no particle
scattering, is not entirely trivial; it has non-trivial collective effects,
and (ii) our renormalized $v$ has real physical significance.

     Since the $h(x)$ field is a non-interacting boson field with mass
$\omega^2 = \frac{1}{2} \lambda_B \phi_B^2 = 8 \pi^2 \phi_R^2$, the only
finite-temperature correction to $V_{{\rm eff}}$ will be the term \cite{dj}:
\begin{equation}
I_1^{\beta} = \frac{1}{\beta} \int \frac{{\rm d}^3 p}{(2 \pi)^3}
\; \ln ( 1- \exp \{- \beta (p^2 + \omega^2)^{1/2} \} ),
\end{equation}
where $\beta = 1/T$, and $T$ is the temperature in units where Boltzmann's
constant is unity.  This term, since it modifies a non-trivial effective
potential, produces non-trivial effects.  In particular, one finds that the
theory undergoes a symmetry-restoring phase transition at a critical
temperature $T_c$ of order $v$.  Note that it is the renormalized $v$, not
$v_B$, that sets the scale for the physical observable $T_c$.  This is
because the {\it depth} of the SSB vacuum relative to the symmetric vacuum
(which is invariant under re-scalings of $\phi$) is of order $v^4$.

    The Gaussian approximation has a very simple generalization to the
finite-temperature case, and after renormalization it yields exactly the
same result as above, because other would-be contributions are infinitesimal
\cite{hajj}.  Thus, we may take over the results from Ref. \cite{hajj},
converting the notation appropriately \cite{fnteft}.  The critical temperature
is $T_c = 0.51394 \, (8 \pi^2/e)^{1/2} \, v = 2.7699 \, v$.  The transition
is first order with a barrier height that is about a quarter of the original
depth of the SSB vacua relative to the origin.  Before the phase transition
the SSB minima of the finite-temperature effective potential move slightly
inward from their zero-temperature positions at $\pm v$.  Because of
this there is a slight (5\%) decrease in the particle mass between $T=0$
and $T = T_c$, with most of this decrease occuring close to $T_c$
\cite{hajj}.

\section{REMARKS ON THE PERTURBATIVE PICTURE}
\setcounter{equation}{0}

    We have insisted on being able to take the continuum limit.  However, if
one is prepared to keep the ultraviolet cutoff finite, so that a finite
``$\lambda_R$'' exists, then one can proceed perturbatively, and here we make
contact with the classic analysis of Coleman and Weinberg \cite{CW}.
Their perturbative
renormalization $\lambda_B = \lambda_R (1 + {\cal O}(\lambda_R))$ leads to
{\it perturbative} expressions for $\beta$ and $\gamma$ \cite{CW}:
\begin{equation}
\label{e27}
 \beta_{{\rm pert}}= + {{3\lambda^2_B}\over{16\pi^2}}~~~~~~~~~~
{\rm and}~~~~~~~~~~\gamma_{{\rm pert}} = O(\lambda^2_B),
\end{equation}
which allow a perturbative solution of the RG equation (\ref{e7}), for
finite ultraviolet cutoff,  valid only up
to higher order terms in $\lambda_R$.  At one-loop order there is then no
wavefunction renormalization, so that $h_B=h_R$ and $\phi_B=\phi_R$.
Making an ``RG improvement'' using $\beta_{{\rm pert}}$ produces a
re-summation of the ``leading log'' series in $x$, defined as
\begin{equation}
\label{e28}
  x={{3\lambda_B}\over{32\pi^2}}\,\ln{{2\Lambda^2}\over{\lambda_B\phi^2_B}},
\end{equation}
yielding a perturbative, running coupling constant:
\begin{equation}
\label{e29}
\lambda_R(\phi^2_R) = {{\lambda_B}\over
{1 + { {3\lambda_B}\over{32\pi^2} }\ln{ {\Lambda^2} \over {\phi^2_R} } }}.
\end{equation}
(At this level of approximation one may drop the $\lambda_B/2$ factor in the
logarithm; it makes only a sub-leading-log difference.)  The resulting
``leading-log effective potential'' has no SSB minima.  However, this
leading-log re-summation procedure is questionable and by no means unique.
The ``leading-log'' series has the form $1-x+x^2-x^3+...$, and converges to
${{1}\over{1+x}}$ only for $|x|<1$.   However, the SSB minimum of the original
one-loop effective potential was precisely at $x = 1$, so it is not surprising
that it can spuriously be made to disappear if one extends the re-summed
expression into the region $x \ge 1$.

 Still, one could argue that the re-summation is trustworthy in the region
where $\phi_B$ is not too small and {\it define} the ``leading-log effective
potential'' over the whole range of $\phi_B$ by analytical continuation.
One would start with a very small bare coupling constant
${{\lambda_B}\over{16\pi^2}} \ll 1$ and a very large cutoff $\Lambda$ such that
$x\sim 1$ but $\lambda_B x \ll 1$ so that the important effects are restricted
to the leading-log sector.  Then, employing $\beta_{{\rm pert}}$ and
$\gamma_{{\rm pert}}$ one would find that all the leading logs are contained
in the running coupling constant $\lambda_R(\phi^2_R)$.  Consequently, one
would
end up with a perturbatively renormalized ``leading log effective action''
that is just the shifted ($\Phi \to \phi + h(x)$) classical action,
with $\lambda_B$ replaced by $\lambda_R(\phi^2_R)$.   However, if one tried
to take the cutoff to infinity with $\beta_{{\rm pert}}$  governing
$\lambda_B$ as a function of $\Lambda$, one would inevitably be drawn into a
region where $\lambda_B$ is {\it not} small.  Then $\lambda_R(\phi^2_R)$
would be driven to zero at any finite $\phi_R^2$.  Alternatively, we can say
that the only region with non-vanishing $\lambda_R(\phi^2_R)$ in which we can
trust the leading-log re-summation is at large $\phi_R^2$, of order
$\Lambda^2$.  In any finite range of $\phi_R^2$ one cannot justify neglecting
``sub-leading'' logarithms.

    The crux of the matter is the qualitative conflict between the one-loop
effective potential itself and its ``re-summed'' or ``RG-improved'' form.
This unhappy situation can be avoided by going to theories such as scalar
electrodynamics.  There, as Coleman and Weinberg argue, there is no doubt
that the leading-log re-summation is valid, for sufficiently small $\lambda$
and gauge coupling $e$.  Our type of analysis, though seemingly very
different, actually leads to the same physical consequences in this
perturbative region \cite{iban,u1}.  However, outside the perturbative
region, the relation between $e_B^2$ and $\lambda_B$ necessary for
renormalizability ceases to be of the form $e_B^4$ proportional to $\lambda_B$.
In fact, $e_B^2$ goes back to zero as $\lambda_B$ approaches the
pure-$\lambda \Phi^4$-theory value, Eq. ({\ref{e15}).  Thus, for small $e_B$
our approach yields two solutions; one perturbative, and one close to pure
$\lambda \Phi^4$ \cite{iban}.  The perturbative solution yields the 10 GeV
Higgs of Coleman and Weinberg, now excluded by experiment.

\section{IMPLICATIONS FOR ELECTROWEAK THEORY}
\setcounter{equation}{0}

     We now consider the implications of this picture for the standard
model of electroweak interactions.  This uses four scalar fields in a
complex isodoublet:
\begin{equation}
\label{K}
K(x)= {{1}\over{\sqrt 2}} ( \chi_1(x) + i\chi_2(x), \, v+h(x)+i\chi_4(x) ),
\end{equation}
where one field $\chi_3(x) = v+h(x)$ has a non-zero vacuum expectation
value.  When this form is substituted into the tree-level scalar-vector
couplings, it generates mass terms for the gauge fields.  Hence, there is
a direct relation between $v$ and the Fermi constant $G_F$; namely,
$v \sim ({\sqrt 2 G_F})^{-1/2} \sim 246$ GeV.  Thus, $v$ is a physical,
measurable quantity, and represents the phenomenological vacuum value of
the scalar field.  The physical origin of its non-zero value, however,
is ascribed to a presently untested part of the theory, namely the ``Higgs
potential''.  In textbook treatments the Higgs potential is treated
classically and there is no `bare/renormalized' distiction.  However, in our
approach the SSB arises only after the full quantum-dynamical content of the
scalar sector has been taken into account.  Therefore, in our approach,
$K(x)$ represents the O(4) extension of our {\it renormalized} field
$\Phi_R(x)$ (i.e. the Fourier transform of $\tilde{\Phi}_R(p)$ introduced
in Sect. 5).  This equals $v_R + h(x)$ when evaluated at the minimum of the
effective potential.  The $v$ in Eq. (\ref{K}) is thus to be identified with
$v_R$, the {\it renormalized} vacuum expectation value, {\it i.e.,} the one
which includes the full dynamical content of the scalar sector.
With $v= v_R$ identified with $(\sqrt{2} G_F)^{-1/2} \approx 246$ GeV our
result $m_h^2 = 8 \pi^2 v^2$ gives a Higgs mass prediction of 2.19 TeV.

    Although the scalar sector must be analyzed and renormalized
nonperturbatively, one may continue to treat the gauge and Yukawa sectors
perturbatively; that is $g_R = g_B(1+O(g^2_B))$.  Hence, one obtains
(up to small corrections from $\mu$-decay) the renormalized relation
$M^2_W=\frac{1}{4} g^2_R v_R^2$, where $g_R$ is the running SU(2) coupling
constant evaluated at the $W$ mass scale.  In the presence of the gauge and
Yukawa couplings, there will be radiative corrections to the effective
potential and to the Higgs mass, but these are small provided the top mass
is less than 200 GeV \cite{con,new}.

    [Note that this attitude is not quite the same as in Refs. \cite{iban,u1}.
In those papers, which deal only with the effective potential, the gauge
coupling was also renormalized in a nonperturbative manner.  We believe that
there is ultimately a duality between the two approaches.  However, for
practical purposes -- since we know that perturbation theory works well in
QED and weak interactions -- the present attitude is more useful.]

    Our mass prediction $m_h^2 = 8 \pi^2 v^2 \approx 2.19$ TeV comes from
considering $\lambda \Phi^4$ theory for a single scalar field, whereas
the standard model involves the O(4) generalization.  In the O(4)
$\lambda \Phi^4$ case the one-loop and Gaussian results are not quite
identical: one-loop gives $m_h = 1.89$ TeV while the Gaussian approximation
gives $m_h = 2.05$ TeV \cite{iban,u1}.  However, this difference is probably
attributable to an inexactness inherent in using ``Cartesian-coordinate''
fields.  Really the Goldstone fields should be described by angular,
``polar-coordinate'' fields.  One can argue that the exact result in the
O($N$) case should be just the $N=1$ result \cite{con,new}; {\it i.e.,} that
only the radial field affects the shape of the effective potential.  This idea
is motivated by the
approach of Ref. \cite{dolan} where the functional integral is expressed in
polar coordinates; the angular integration gives only a constant term and the
non-trivial Jacobian is handled by ghost fields \cite{fnoteon}.

   In any case, one predicts a Higgs mass $m_h \sim $ 2 TeV in the Standard
Model.  Moreover, despite its large mass, the Higgs would be narrow, decaying
principally to $t \bar{t}$, and there would be no strong interactions among
longitudinal gauge bosons.  This follows, by the ``equivalence theorem''
(see below), from the fact that the scalar sector is non-interacting in the
absence of gauge couplings.  Now, it might seem at first that ``triviality''
for broken-phase O$(N)$ theory would be at odds with current algebra, which
prescribes definite derivative couplings for Goldstone bosons.  However,
this conflict can be reconciled by the fact that $v_B$ is infinite, which
causes the current-algebra interactions in O$(N)$ $\lambda \Phi^4$ theory to
be infinitely suppressed.   Thus, both sides of an ``Adler-Weisberger'' sum
rule would be infinitesimal: the scattering cross sections are infinitesimal,
reflecting ``triviality'', while the other side of the equation is supressed
by a $1/v_B^2$ factor.  The scalar self-interactions disappear because $v_B$
is infinite, but it is the finite $v_R$ that sets the scale of the symmetry
breaking, and governs $T_c$ and the particle masses.

    Our picture is perfectly compatible with the ``equivalence theorem''
\cite{eqth,bagger}, whose physical content, paraphrasing Sect. II of
Ref. \cite{bagger}, is the following:  For zero gauge coupling(s), $g$, the
Goldstone bosons are physical particles while the longitudinal components of
the $W$'s are free, unphysical degrees of freedom, included just to maintain
manifest Lorentz covariance.  We call this the ``$g=0$ theory''.   On the
other hand, for $g\neq 0$, however small, the longitudinal $W$'s are physical
while the Goldstone bosons are now unphysical particles, included just to
preserve renormalizability.  This situation we call the ``$g \to 0$ theory''.
The requirement that the physical observables of the two theories are the same
in the limit $g\to 0$ implies an equivalence between the physical longitudinal
$W$'s of the ``$g \to 0$ theory'' and the physical Goldstone bosons of the
``$g=0$ theory.''  This statement is made precise by the theorem, which is
valid to lowest non-trivial order in $g$ and to {\it all orders} in the scalar
self-interaction \cite{eqth,bagger}.  Thus, presumably the theorem remains
valid for a nonperturbative scalar sector.  In our
picture the $g=0$ theory has non-interacting Higgs and Goldstone particles,
so in the small-$g$ theory we expect only weakly interacting Higgs and
longitudinal gauge bosons.  So, at low energy scales, the only trace of the
Higgs particle is represented by the logarithmic one-loop correction to
the $\rho$ parameter discovered by Veltman \cite{velt}.

   In our picture there is just no such thing as a ``renormalized $\lambda$'',
and the quasi-classical relation ``$m_h^2 \propto \lambda_R v^2$'', implying
a proportionality between the Higgs mass and its self-coupling, is completely
misleading.  As has been pointed out by Huang \cite{huang2,huang}, the ratio
$m_h^2/v^2$ is not a measure of the effective scalar coupling strength.

\vskip 10 pt
\centerline {ACKNOWLEDGEMENTS}
\par One of us (M.C.) would like to thank A. Agodi, R. Akhoury, M. Einhorn,
P. Federbush, K. Huang, R. Jackiw, G. Kane, Y. Tomozawa, M. Veltman and
Y. Yao for very useful discussions.
This work was supported in part by the U.S. Department of Energy under
Grant No. DE-FG05-92ER40717.

\vfill
\eject


\begin{thebibliography} {99}
\bibitem{fnote}
We do not consider negative $\lambda_B$, which is a whole other story.
\bibitem{fnteft}
The point that $T_c$ is proportional to the renormalized $v$ was noted in
Ref. \cite{iban}, but the numerical value was quoted incorrectly.
\bibitem{fnoteon}
Note that the range of
integration for the shifted radial field, $- v_B$ to $+ \infty$, will
effectively become $- \infty$ to $+ \infty$ because of the infinite $Z_{\phi}$
re-scaling.  Similarly, the angular fields, defined on $-\pi$ to $\pi$, would
become normal massless scalar fields, defined on $- \infty$ to $+ \infty$ when
re-scaled by $v_B$.
\end{thebibliography}
\end{document}